\documentclass[amsmath,amssymb,12pt]{revtex4}

\usepackage{diagrams}

\newcommand{\dpz}{\Delta\dot\phi_0}
\newcommand{\dpe}{\Delta\dot\phi_E}
\newcommand{\state}[1]{\vert #1\rangle}
\newcommand{\hil}{H}
\newcommand{\htot}{\hil_{\text{\scriptsize tot}}}
\newcommand{\hsys}{\hil_{\text{\scriptsize sys}}}
\newcommand{\us}{U_{\text{\scriptsize sys}}}
\newcommand{\happ}{\hil_{\text{\scriptsize app}}}
\newcommand{\uapp}{U_{\text{\scriptsize app}}}
\newcommand{\henv}{\hil_{\text{\scriptsize env}}}
\newcommand{\uenv}{U_{\text{\scriptsize env}}}
\newcommand{\hrest}{\hil_{\text{\scriptsize rest}}}

\newcommand{\psis}{\psi_{\text{s}}}
\newcommand{\R}{\mathbb{R}}

\newcommand{\ham}{\text{\textsf{H}}}
\newcommand{\ens}{\mathcal{E}}
\newcommand{\bsigma}{\mbox{\boldmath $\sigma$}}
\newcommand{\bham}{\mbox{\boldmath $\ham$}}
\newcommand{\rd}{\textbf{RD}}
\newcommand{\tsb}{T_{\text{\scriptsize SB}}}

\begin{document}
\title{Emergent Probabilities in Quantum Mechanics}
\author{Olaf Dreyer}
    \email{o.dreyer@imperial.ac.uk}
    \affiliation{Theoretical Physics, Blackett Laboratory, 
    Imperial College of Science, Technology and Medicine,
Prince Consort Road, London, SW7 2AZ, U.K.}
\date{March 22, 2006}
\begin{abstract} The transition from the quantum to the classical is governed by randomizing devices (RD), i.e., dynamical systems that are very sensitive to the environment. We show that, in the presence of RDs,  the usual arguments based on the linearity of quantum mechanics that lead to the measurement problem do not apply. RDs are the source of probabilities in quantum mechanics. Hence, the reason for probabilities in quantum mechanics is the same as the reason for probabilities in other parts of physics, namely our ignorance of the state of the environment. This should not be confused with decoherence. The environment here plays several, equally important roles:  it is the dump for energy and entropy of the RD, it puts the RD close to its transition point and it is the reason for probabilities in quantum mechanics. We show that, even though the state of the environment is unknown, the probabilities can be calculated and are given by the Born rule. We then discuss what this view of quantum mechanics means for the search of a quantum theory of gravity.
\end{abstract}

\maketitle
\section{Introduction}
One of the most striking features of quantum mechanics is the probabilistic nature of its predictions.   It is generally believed that this is a fundamentally new feature of the quantum world.  This is in contrast to the way usually probabilities arise in our description of nature.   Before quantum mechanics, probabilities arose because we faced situations where we had incomplete knowledge of the state of the system. In this article shall argue that in fact the probabilities in quantum mechanics are of this exact same type.  

We shall show that the key to understanding the probabilistic nature of quantum mechanics is what we call a randomizing device.  Any measurement involves a randomizing device, the measurement apparatus.  This maps the states of a small quantum system to the states of a large quantum system.  It is the emergent properties of the large quantum system that constitute what we call the measurement outcome.  In the process of emergence the randomizing device is very sensitive to the environment and it is here that probabilities enter.  

The paper is organized as follows. In the next two sections we argue that \rd's are common and give examples of them. We further point out the new roles the environment plays in this view of quantum mechanics. In section \ref{sec:born} we show that it is the Born rule \cite{born} that gives the probabilities for the measurement outcomes. In the last section \ref{sec:outlook} we review the paper and discuss its meaning for the search of a quantum theory of gravity.

\section{Randomizing Devices}
Some of the key issues that arise in the measurement problem can be illustrated using a simple example from classical mechanics, a pendulum.

Let $\phi$ be the angle of the pendulum to the vertical and let the mass, the length, and the gravitational constant $g$  all be unity. The equation of motion for the pendulum is
\begin{equation}
\ddot\phi + \sin \phi = 0.
\end{equation}

When the initial position of the pendulum is $\phi_0=0$ it is hanging straight down. We are interested in the special case when the initial angular velocity of the pendulum is such that it {\em just} reaches the top. This happens when the initial angular velocity $\dot\phi_0$ is
\begin{equation}
\frac{1}{2}\dot\phi_0^2 -1 =1,
\end{equation}
or,
\begin{equation}
\dot\phi_0 = 2.
\end{equation}
With this angular velocity, the pendulum will just make it to the top and will require an infinite amount of time to do it. Now consider angular velocities in the vicinity of $\dot\phi_0 =2$, i.e.,
\begin{equation}
\dot\phi_0 = 2 + \dpz.
\end{equation}
For $\dpz>0$ the pendulum will make it over the top and for $\dpz<0$ the pendulum will swing back before making it to the top. Let the initial push of the pendulum come from the right and denote the outcome where the pendulum reaches the top and then swings further $R$ and the outcome where it does not reach the top $L$. Thus, $\dpz>0$ is outcome $R$ and $\dpz<0$ is outcome $L$.

Now note that this analysis is not the realistic setup of a pendulum with an environment.  For $\dpz$ small enough, any fluctuation of the environment will influence the outcome of the experiment. So when
$\dpz > 0$,
all we can conclude is that it is more likely that outcome $R$ will occur. For $\dpz$ small enough  the pendulum might also end up in outcome $L$. 

Let us try to quantify the effect of the environment. We will assume that the effect of the environment is described by just one contribution, $\dpe$,  to $\dot\phi_0$.  The effective $\dot\phi_0$ will thus be
\begin{equation}
\dot\phi_0 = 2 + \dpz + \dpe.
\end{equation}
The likelihood for $\dpe$ actually occurring is described by a probability distribution 
$p_E( \dpe)$. What are then the probabilities for the outcomes $L$ and $R$? To obtain outcome $R$ we need
\begin{equation}
\dpz + \dpe > 0.
\end{equation}
The probability for this to happen is
\begin{equation}
p_R = \int_{-\dpz}^\infty dx\,  p_E(x)\ > \ 0.
\end{equation}
Similarly,  the probability for $L$ occurring is
\begin{equation}
p_L = \int_{-\infty}^{-\dpz} dx\,  p_E(x) \ > \ 0.
\end{equation}

Now we will stretch the notation somewhat to make the connection to the quantum mechanical situation clearer. When the initial angular velocity of the pendulum is $2 + \dpz$, we will say that the pendulum is in the state $\vert \dpz \rangle$. When the outcome is $L$ ($R$) we will say that the state of the system is 
$\vert L\rangle\ (\vert R\rangle)$.
We will also introduce the amplitudes
\begin{equation}
q_{L,R} = \sqrt{p_{L,R}}
\end{equation}
for the probabilities $p_{L,R}$ calculated above. 
Then we can write schematically
\begin{equation}
\vert \dpz \rangle = q_L \vert L\rangle + q_R \vert R\rangle.
\end{equation}
The coefficients of the states $\vert L\rangle$ and $\vert R\rangle$ are the probabilities of the corresponding outcomes as would be the case in a  quantum mechanical system.

We do not want to stretch the analogy too far but there are a couple of points that this example illustrates. The most important point the pendulum example makes is to reveal the enormous role played by the environment when such a dynamical system  at a critical point is present. In our classical example this role is played by the pendulum that is given a push that just about balances it on the top. A very small deviation from the initial push will decide whether the system will swing to the right or the left. For future reference let us call such a dynamic system a {\em randomizing device} (\rd). We will argue that quantum mechanical {\em measurement devices are always \rd's}. 

The next thing to note is that before the experiment with the pendulum (feel free to substitute here measurement for experiment) is made there is no sense in which the system is in outcome $L$ or $R$. Only after the experiment is done can we talk about the outcome. This will hold true also in the quantum case. 

\section{The measurement problem in the presence of a randomizing device}\label{sec:rd}
A typical example of a \rd\ in quantum mechanics is a many-particle system that is about to undergo a symmetry breaking transition. As an example, we shall consider the one-dimensional Heisenberg model. It is described by the Hamiltonian 
\begin{equation}
  \bham = \sum_{l=1}^N \bsigma_l\cdot\bsigma_{l+1}\label{eq:scham}.
\end{equation}
Above a certain temperature $\tsb$, this system is in a symmetric state with the symmetry group SU(2). Below $\tsb$, the system undergoes a symmetry breaking transition to one of its ground states.  Near $\tsb$, the system is very sensitive to the state of the environment. By tuning the the temperature to a value close to the transition temperature, i.e., by choosing
\begin{equation}
T = \tsb + \epsilon, 
\end{equation}
with $\epsilon>0$, one can make the system as sensitive as desired. If $\epsilon$ is small enough the system can be used to measure the state of a single spin $\bsigma_0$ by coupling it to the spin chain. 

A measurement device needs to be a randomizing device.   A measurement apparatus is set up to be in a delicate enough state that the system to be measured can easily push it to one of the outcome states. A spin chain can be used as a measurement apparatus. A cloud chamber is another example, of a system brought into a state close to the point where the gas-liquid transition occurs.  Here the environment provides the pressure and the temperature to hold the chamber at a point where it is very sensitive to outside perturbations.

One may object that the environment should be in a symmetric state that does not favor any one ground state. This is true, but one has to realize that the environment is in a symmetric state {\em only in an ergodic sense}. At any given moment it will push the chain towards one of the ground states. Only in a time averaged way is it symmetric: the chain is pushed towards each ground state an equal amount of the time. 

In classical mechanics, the fact that the time evolution depends crucially on the initial conditions has been investigated for some time. It is the subject of chaos theory. The view of a \rd\ offered here shares with classical chaos the sensitivity to initial conditions but it goes beyond this. In chaos theory the system is always described by the same set of variables like position and momentum. It is just not known what the values of  these variables is. An \rd\ does more in that it produces in the measurement process a state with qualitatively new properties. In the chain considered above these new properties are order and rigidity. Both of these properties can not be formulated on the level of a single spin. Thus \rd's are more powerful then chaos since they not only introduce randomness they also produce genuine novelty. It is in the context of quantum mechanics that more is really different \cite{anderson}.

\subsection{The measurement problem}

Let us now look at the measurement problem given that the measurement apparatus is a randomizing device. 
The measurement problem in quantum mechanics arises because of its linear structure. If a system is in a state $\state{a}$ ($\state{b}$) and the measurement apparatus ends up being in the state $\state{A}$ ($\state{B}$) then it follows that if the system is in a linear superposition $
\alpha\state{a} + \beta\state{b} $, the outcome of the measurement should be 
\begin{equation}\label{eqn:super}
\alpha\state{A} + \beta\state{B}.
\end{equation}
The measurement problem is that no such superposition has ever been observed. 

The logic of this argument is flawed when a \rd\ is present. In such a case the environment can not be neglected. The probabilistic nature of the environment will destroy the linearity assumed in the argument above. Let us thus repeat the above argument, this time including the environment. Beginning with the state of the system being $\state{a}$, we have that 
\begin{equation}
\state{a}\state{N}\state{e_1}
\end{equation}
evolves into 
\begin{equation}\label{eqn:measa}
\state{a}\state{A}\state{e_1^\prime},
\end{equation}
where $\state{N}$ denotes the neutral state of the apparatus. Similarly for $\state{b}$ we have
\begin{equation}\label{eqn:measb}
\state{b}\state{N}\state{e_2} \longrightarrow \state{b}\state{B}\state{e_2^\prime}.
\end{equation}

It is key that the state of the environment in the second repeat is different than its state in the first run.  This contrasts the common assumption that takes the environment to be in the same state on the second run (here $\state{e_1}$) as before. This is clearly wrong.  With every new measurement the environment is in a new state. However carefully the measurement is prepared this fact does not change. The true situation then is
\begin{equation}
(\alpha\state{A} + \beta\state{B})\state{N}\state{e_3}.
\end{equation}
What can we deduce about the evolution of this superposition given what we know from equations (\ref{eqn:measa}) and (\ref{eqn:measb})? The answer is not much. We can only arrive at (\ref{eqn:super}) if we neglect the influence of the environment on the apparatus.  Since the measurement apparatus is an \rd\ this is exactly what we can not do. Even though the theory is fundamentally linear, equation (\ref{eqn:super}) does not follow. 

We propose that it is this fact that provides a solution to measurement problem. 
If one takes into account the role of \rd's then it is no longer enough to point to a state of the form $\state{A} + \state{B}$ in the Hilbert space and say there is a problem. Instead one has to show how dynamically such a state could arise. In the presence of an \rd\ this is very hard.
It is also here that probabilities enter quantum mechanics. Probabilities in quantum mechanics have the same status as probabilities elsewhere in physics. They arise because of our incomplete knowledge. No fundamental dice are needed. 

\subsection{The many roles of the environment}
It has been argued elsewhere that classical states are to be identified with symmetry broken states of large quantum systems \cite{anderson2, laughlin}.  Here we argued that a symmetry breaking transition is a prime example for what we have called a \rd, a randomizing device. If classical states can only be reached by a transition of this type then the quantum-to-classical transition is by necessity a random one. This is the main contention of this paper. 

In the literature, the role of the environment on quantum mechanics is usually restricted to decoherence (see \cite{zurek1} and references therein). Here we saw the environment play a number of other important roles that are not commonly acknowledged:
                               
\begin{enumerate}
\item The environment has to bring the \rd\  close to a transition point. Near this point the apparatus is sensitive to the state of the system but also to the environment. 

\item The environment  is a dump for energy and entropy for the apparatus. The state of the apparatus before the measurement, i.e. the neutral state, is one of higher entropy and energy than the state of the apparatus after the measurement. The environment is there to absorb the difference. An important consequence of this is that without an environment there is no measurement. 

\item Through the coupling of the environment to the apparatus an irreducible element of chance is introduced. It is the presence of the environment that gives the outcomes their probabilistic nature. In the next section we show that although we do not know the state of the environment we still can calculate the probabilities of outcomes. 
\end{enumerate}

\section{The Born Rule}\label{sec:born}
In the last section we saw that a randomizing device is an essential element of the measurement process.  With the randomizing device there comes an irreducible element of chance. What then are the corresponding probabilities? In this section we want to calculate the probabilities of measurement outcomes and show that they coincide with the probabilities given by the Born rule \cite{born}. For this we will make use of arguments first introduced by D. Deutsch \cite{deutsch} and D. Wallace \cite{wallace} in the context of the many worlds interpretation of quantum mechanics. Later S. Saunders \cite{saunders} stripped the arguments of their many worlds baggage and more recently W. Zurek \cite{zurek2} used the same arguments with yet another motivation.

In a measurement process, there are  three parts, the system Hilbert space $\hsys$, the apparatus Hilbert space $\happ$ and the environment Hilbert space $\henv$, making up the total Hilbert space $\htot$:
\begin{equation}
\htot = \hsys \otimes \happ \otimes \henv.
\end{equation}
 Of special importance is the apparatus because it is the randomizing device. Sometimes it will be convenient not to distinguish between the environment and the apparatus. Both are large quantum systems and their precise state at the beginning of the measurement is not known to us. This is why we will often treat them together:
\begin{equation}
\hrest = \happ \otimes \henv.
\end{equation}

Let $I$ denote the set of possible measurement outcomes. Since the apparatus and the environment are large systems there will be a large number of states that correspond to the same measurement outcome. Let $O$ be those states in $\htot$ that correspond to measurement outcomes and let 
\begin{equation}
a : O \longrightarrow I
\end{equation}
be the map that maps a state in $O$ to its corresponding measurement outcome.

The aim of this section is to calculate the probability $p_i(\psis)$ for a given outcome $i\in I$ given a state $\psis\in\hsys$.
It would seem that there is very little that constrains the probabilities $p_i$. We shall see that there are a number of constraints on the $p_i$'s that allow us to  calculate the probabilities.

\subsection{Symmetries}

Since we do not know the state of the environment and the apparatus at the beginning of the measurement, we have to use an ensemble $\ens$ of states in $\hrest$ to describe the possible states. We shall not be too picky about which states to include in the ensemble $\ens$. We shall ask  for just one thing:  if the Hamiltonian of the system and the apparatus has a symmetry,  the ensemble must respect this symmetry. By this we mean that if $U = \us\otimes\uapp$ is a unitary implementing the symmetry on $\hsys\otimes\happ$ then we assume that there is an extension $\tilde U$ to the whole Hilbert space of the form 
\begin{equation}
\tilde U = \us\otimes\uapp\otimes \uenv
\end{equation}
and that  the ensemble $\ens$ is such that
\begin{equation}
\chi\in\ens \text{\ \ if and only if\ \ } (\uapp\otimes\uenv)\chi\in\ens.
\end{equation}
That is, the ensemble is symmetric under the same symmetries as the Hamiltonian of the system and the apparatus. This is a natural assumption to make for otherwise a symmetric Hamiltonian will not lead to a symmetric evolution. 

Now that we have narrowed down the type of ensemble $\ens$, let us take a closer look at the kind of symmetries we are interested in. Let 
\begin{equation}
U = \us \otimes \uapp
\end{equation}
be as above and let $\tilde U$ be the extension of $U$ to the whole Hilbert space $\htot$.  We assume that $\tilde U$ commutes with the total Hamiltonian $\ham$ 
\begin{equation}
[ \tilde U, \ham ] = 0.
\end{equation}
We are especially interested in those $\tilde U$'s that map measurements into measurements. We thus want $\tilde U$ to be such that there exists a map
\begin{equation}
\iota : I \longrightarrow I,
\end{equation}
so that the diagram
\begin{diagram}
    O & \rTo^a & I & \\
    \dTo^{\tilde U} & & \dTo_{\iota} \\
    O & \rTo^a & I
\end{diagram}
commutes.

If such a symmetry $\tilde U$ exists, we can derive a rule that the $p_i(\psis)$'s have to satisfy. Given the ensemble $\ens$ the $p_i(\psis)$'s are proportional to the number of states $\chi\in\ens$ for which
\begin{equation}\label{eqn:calcp}
a( U_T (\psis\otimes\chi) ) = i,
\end{equation}
where $U_T = \exp{(i \bham T)}$ is the time evolution for some time interval $T$ larger than the time required to perform the measurement. 

Now let $\chi\in\ens$ be such that (\ref{eqn:calcp}) is true.  It follows that
\begin{eqnarray}
a( U_T \tilde U  (\psis\otimes\chi) ) & = & a( \tilde U U_T  (\psis\otimes\chi) ) \\
 & = & \iota( i).
\end{eqnarray}
This means that the number of states $\chi\in\ens$ for which the measurement gives $i\in I$, given $\psis\in\hsys$, is the same as the number of states $\chi\in\ens$ for which the measurement will give $\iota(i)\in I$ given $\us\psis\in\hsys$. We have thus shown that if $U$ is as above we have
\begin{equation}\label{eqn:rule}
p_i(\psis) = p_{\iota(i)}(\us \psis).
\end{equation}

\subsection{From symmetries to the Born rule}

Having established a general rule (\ref{eqn:rule}) that the probabilities $p_i(\psis)$ have to satisfy we now want to look at some particular implementations of this rule. We will let the spin chain Hamiltonian of eq.(\ref{eq:scham})
guide us in our argumentation. Let spin $\bsigma_0$  be  the system
and the rest of the spin chain be the apparatus. The Hamiltonian is symmetric under SU(2) rotations. 
Acting on the spin chain with an  element of SU(2), for example,
\begin{equation}
U_\phi = \begin{pmatrix}
e^{i\phi} & 0\\
0 & e^{-i\phi}
\end{pmatrix}\in \text{\ SU(2)}, 
\end{equation}
does not change the orientation of the spins. The map $\iota: I \longrightarrow I$ is thus the identity:
\begin{equation}
\iota = \text{id}_{I}.
\end{equation}
We will assume that such a $U_\phi$ always exist. It then follows that the $p_i$'s have to satisfy Property 1:

\begin{description}
\item[P1] For all $\psis\in\hsys$ and all $i\in I$,
\begin{equation}
p_i(\psis) = p_i(U_\phi \psis),
\end{equation}
where $U_\phi$ is given by
\begin{equation}
U_\phi = \text{diag}( 1, \cdots, 1, e^{i\phi}, 1, \cdots, 1, e^{-i\phi}, 1, \cdots, 1).
\end{equation}
\end{description}

The next symmetry we want to look at is the exchange of up and down. For 
\begin{equation}\label{eqn:upi}
U_\pi = \begin{pmatrix}
0 & 1 \\
1 & 0
\end{pmatrix}
\end{equation}
we have
\begin{eqnarray}
U_\pi \sigma_z U_\pi & = & -\sigma_z \\
U_\pi \sigma_x U_\pi & = & \sigma_x \\
U_\pi \sigma_y U_\pi & = & -\sigma_y.
\end{eqnarray}
Since $\bham$ is quadratic in the $\sigma$'s the exchange given by $U_\pi$ is a symmetry of the Hamiltonian. In this case, the map $\iota$ is 
\begin{equation}
\iota = \pi,
\end{equation}
i.e. the exchange of up and down. Again we assume that such a transformation is generically present. This leads to Property 2  for the $p_i$'s:

\begin{description}
\item[P2] For all $\psis\in\hsys$ and all $i\in I$ we have
\begin{equation}
p_i(\psis) = p_{\pi(i)} (U_\pi\psis),
\end{equation}
where $\pi$ is a permutation of the elements of $I$ and $U_\pi$ is the representation of $\pi$ on $\hsys$. For a two dimensional $\hsys$ it $U_\pi$ is given by (\ref{eqn:upi}). 
\end{description}

It is surprising that from these two properties we can already calculate the probabilities for the case of amplitudes of equal magnitude. For the sake of notation we will concentrate on the case of two dimensions. Let 
\begin{equation}
\psis = \alpha \state{a} + \beta\state{b},
\end{equation}
with $\vert\alpha\vert = \vert\beta\vert = 1/\sqrt{2}$. Properties \textbf{P1} and \textbf{P2} in this case allow for two different ways to exchange the amplitudes $\alpha$ and $\beta$. \textbf{P1} does not affect the measurement outcome whereas \textbf{P2} does. 

In the following calculation we first use \textbf{P1} with $\phi\in\R$ such that
\begin{equation}
e^{i\phi} = 2\bar\alpha\beta,
\end{equation}
i.e. 
\begin{equation}
U_\phi = 2\begin{pmatrix}
\bar\alpha\beta & 0 \\
0 &  \alpha\bar\beta
\end{pmatrix}.
\end{equation}
After that we use $U_\pi$ as it is given above:
\begin{eqnarray}
p_i( \alpha \state{a} + \beta\state{b} ) & = & p_i( \beta \state{a} + \alpha\state{b} )\\
 & = & p_{\pi(i)}( \beta\state{b} + \alpha \state{a} )
\end{eqnarray}
Thus for all $\psis = \alpha\state{a} + \beta\state{b}$ with $\vert\alpha\vert = \vert\beta\vert$ we have
\begin{equation}
p_i(\psis) = p_{\pi(i)}(\psis).
\end{equation}
Since the sum of the $p_i$'s is unity we have
\begin{equation}
p_i(\psis) = \frac{1}{2}.
\end{equation}

It is clear that the above derivation does not depend on the dimension of $\hsys$. In general we thus have 
\begin{equation}
p_i(\psis) = \frac{1}{n},
\end{equation}
for all
\begin{equation}
\psis = \sum_{j=1}^n\alpha_j\state{j},
\end{equation}
with $\vert\alpha_k\vert = \vert\alpha_l\vert$ for all $k$ and $l$.

The same reasoning can be generalized to the case of unequal amplitudes.  To do so, one adapts the results of \cite{saunders,zurek2} to our setup. 

\section{Conclusion}\label{sec:outlook}

We have seen that the key to understanding the probabilistic nature of quantum mechanics is what we have called a randomizing device.  Any measurement involves a randomizing device, the measurement apparatus.  This maps the states of a small quantum system to the states of a large quantum system.  For example, if we use a spin chain as a measuring apparatus, we identify its up and down states with those of the system and say that the state of the system after a measurement is either up or down.  However, for the spin chain to be a good measurement apparatus, it needs to be at an unstable state that collapses to an up or down state when coupled to the system.  That is, when we perform a measurement we couple it to a randomizing device whose emergent properties are then used to describe the system. Since this process is by necessity probabilistic the whole theory appears probabilistic. This identification of the state of the system with the state of a larger quantum system that has undergone a phase transition or similar collapse from an unstable state is what lies at the heart of the measurement problem of quantum mechanics. 

Central to this process is the environment that the \rd\  is coupled to. The role of the environment in our understanding of quantum mechanics has been so far restricted to decoherence. In this paper we have seen that the environment has more, equally important roles to play:  It is a dump for energy and entropy of the \rd. It puts the \rd\ into a position close to the transition point and finally it is the reason why quantum mechanics is probabilistic. 

Although the state of the environment is unknown the probabilities the environment creates are highly constrained. We have shown that they coincide with the usual Born rule. 

This provides a new answer to the measurement problem of quantum mechanics.  
The states of the apparatus are states of a large quantum system. A large system is required for these states to exist. In the spirit of ``more is different'' \cite{anderson}: We use the emergent properties of a large quantum system to characterize a small quantum system that on its own does not have these properties. Quantum mechanics appears strange because we use expressions based on emergent properties of randomizing devices to describe systems that do not and can not have these properties. In short: Our daily world is one level of emergence away from the quantum world.

In the beginning of the last century theoretical physics faced a severe conceptual problem. The second law of thermodynamics had been identified as one of the pillars of statistical mechanics but there remained the troubling issue of Poincar\'e recurrence. How could the second law of thermodynamics be true if the system was bound to return to a state arbitrarily close to the one it started from? A solution to this problem was given by the Ehrenfests \cite{ehrenfest}: for all practical purposes the system will not return to its initial position. It is just too unlikely. In quantum mechanics, the door to a solution of this kind, i.e., a solution for all practical purposes, was closed by John Bell \cite{bell}. He explicitly introduced  the shorthand FAPP and gave it a bad name. He demanded a ``real'' solution to the measurement problem. In this paper we have argued that {\em only} a solution for all practical purposes exists. When it comes to the measurement problem we are in a situation not unlike the situation faced by statistical mechanics in the beginning of the century.

We have seen that the measurement problem is linked to the dynamical behavior of large quantum systems at a critical point. A deeper understanding of the measurement problem could be achieved by further study of the transition dynamics. This is a problem that is notoriously hard. 

Note also that the discussion in the present paper has a bearing on existing suggestions for the emergence of classicality from a quantum system. We have seen that the outcomes of measurements are determined by the \emph{dynamics} of the system and the apparatus. It is the groundstates of the apparatus that give the measurement outcomes. If we want to know the classical states of the a system we have to solve its dynamics and find its ground states. From this it follows that coherent states are ill suited to describe classical states because they are completely kinematical. 

For a general large quantum system solving the dynamics is a nearly impossible task. It is because this task is so hard that so little progress is being made in theories where a certain quantum dynamics is assumed and the classical limit is looked for. 

Finally, the present view of quantum mechanics is in line with approaches to quantum gravity in which our classical views about space and continuous time are based on emergent properties like extension. That is, it is the emergent property of rigidity that is largely responsible for our notion of space. The present work has implications for such a quantum theory of gravity.  Just as discussed above, the emergent properties of a large quantum system can be used to characterize a small quantum system that on its own does not have these properties (such as extension or rigidity).  Hence the fundamental theory should not be based on objects having properties of spacetime geometry, even in a quantum form, since geometry has to be based on the emergent properties. This also implies that the fundamental theory can not be obtained by a process of quantization. 

\begin{acknowledgements}
The author would like to thank Lee Smolin, Chris Beetle, Chris Isham, Fay Dowker, Fotini Markopoulou, Hans Westman, Joy Christian and the members of the Perimeter Institute for their time and comments.
\end{acknowledgements}


\begin{thebibliography}{99}
\bibitem{born} M.~Born, Z. Phys. \textbf{37}, 863 (1926).
\bibitem{anderson} P.~W.~Anderson, Sience \textbf{177}, 393 (1972).
\bibitem{anderson2} P.~W.~Anderson, \emph{Basic Notions of Condensed Matter Physics} (Westview Press, 1997).
\bibitem{laughlin}  R.~B.~Laughlin, \emph{A Different Universe: Reinventing Physics from the Bottom Down} (Basic Books, 2005).
\bibitem{zurek1} W.~H.~Zurek, Physics Today \textbf{44}, 36 (1991). An updated version of this article is available at quant-ph/0306072. 
\bibitem{deutsch} D.~Deutsch, Proc. R. Soc. Lond. \textbf{A455}, 3129 (1999).
\bibitem{wallace} D.~Wallace, Stud. Hist. Phil. Mod. Phys. \textbf{34}, 415 (2003).
\bibitem{saunders} S.~Saunders, Proc. R. Soc. Lond. \textbf{A460}, 1 (2004).
\bibitem{zurek2} W.~H.~Zurek, Phys. Rev. Lett. \textbf{90}, 120404 (2003).
\bibitem{ehrenfest} P.~Ehrenfest and T.~Ehrenfest, \emph{The Conceptual Foundations of the Statistical Approach in Mechanics} (Dover, 2002).
\bibitem{bell} J.~S.~Bell, Phys.World \textbf{3}, 33 (1990). This article is reprinted in J.~S.~Bell, \emph{Speakable and Unspeakable in Quantum Mechanics}, (Cambridge University Press, 2004).  
\end{thebibliography}
\end{document}